# A Proposed Practical Problem-Solving Framework for Multi-Stakeholder Initiatives in Socio-Ecological Systems Based on a Model of the Human Cognitive Problem-Solving Process

Kevin Kells

University of Ottawa, Telfer School of Management, Kevin.Kells@uottawa.ca

https://telfer.uottawa.ca/en/directory/professors/kells-kevin

A practical problem-solving framework is proposed for multi-stakeholder initiative (MSI) problem-solving processes involving socio-ecological systems (SES)—so-called wicked problems—based on insights borrowed from a model of the individual human, cognitive problem-solving process. The disciplined facilitation of the multi-stakeholder process, adhering to the steps recognized in the individual process, is meant to reduce confusion and conflict. Obtaining a one- to three-sentence human-language description of the desired system state, as a first step, is proposed in multi-stakeholder initiatives for reasons of goal congruence and trust building. The systematic, stakeholder-driven subdivision of obstacles into larger numbers of simpler obstacles is proposed in order to obtain a list of "what needs to be done," inviting a more rational and goal-driven conversation with resource providers. Finally, obtaining and maintaining stakeholder buy-in over the course of the problem-solving effort is reinforced by reflecting back to all stakeholders, as a communication device, a dynamic, visual problem-solving model, taking into account the diversity of cognitive and individual capacities within the stakeholder group in its presentation. Mathematical parameters for gauging applicability of the proposed framework are discussed.

## 1 Introduction

The present work aims to contribute to practical tools for approaching humanity's grand challenges, such as climate change, health care, poverty, education quality and equity, crime, corruption, socio-economic disparity, and political polarization. Churchman (1967) and Rittel & Webber (1973) identified such multi-stakeholder problems in socio-ecological systems as "wicked," a classification for difficult to define problems which defy "silver bullet" solutions. Head and Xiang give an overview of the characteristics of these problems (2016) while Crowley and Head provide a modern perspective on the original work nearly a half-century later (2017). Peters (2017) contributes a conceptual analysis about what is "wicked" about wicked problems, while Andersson and Törnberg (2018) develop a comprehensive *meta-*ontological map of these problem types.

Work on solution approaches has provided insight and refinement of strategies (e.g., Buchanan, 2006; Conklin, 2007; Dentoni, Bitzer, & Schouten, 2018; Ferraro & Beunza, 2018; Foley, Wiek, Kay, & Rushforth, 2017; Kusters et al., 2018; Lemon & Pinet, 2018; MacDonald, Clarke, & Huang, 2018; Nowell, 2010; Riley et al., 2015; Roberts, 2000; Scharmer, 2016; Sova et al., 2015; Veldhuis, van Scheepstal, Rouwette, & Logtens, 2015; Voltan & De Fuentes, 2016; Weymouth & Hartz-Karp, 2015).

Problem structuring (Shaw, Westcombe, Hodgkin, & Montibeller, 2004) using cognitive mapping (Ackermann, Cropper, & Eden, 1992; Damart, 2010) and causal mapping (Eden, Ackermann, & Cropper, 1992; Veldhuis et al., 2015) has been proposed and studied, as has the establishment of goals using Group Decision Support Systems (Cil, Alpturk, & Yazgan, 2005; Elia & Margherita, 2018).

Innes and Booher (2016) describe a collaborative practice that they call "collaborative rationality," while Kania et al. (2014) describe their "collective impact" multi-stakeholder efforts. The conditions for successful multi-stakeholder work identified by these authors is summarized as follows:

1. An all-inclusive stakeholder representation, meeting face-to-face with the guidance of skilled, independent facilitation staff.
2. The focus on a problem common to all stakeholders and development of a common vision, joint approach, and agreed-upon actions.
3. A shared confidence in commonly understood and agreed-upon information and measurement of success indicators.
4. A multi-phase stakeholder meeting process with disciplined adherence to each phase.
5. Frequent and structured communication for trust, motivation, and work towards common objectives.

Our contribution is meant to be a pragmatic, action-oriented, and participatory framework. Above all, we are deliberate in applying technology appropriate to enabling and empowering individual human potential, which we think is key in rising to this category of challenges (Brown, 2015).

## 2 Our Proposed Contribution

The proposed framework integrates the following elements.

- We strive to meet the five conditions set out by Innes and Booher (2016) and Kania et al. (2014).
- We propose a stakeholder-driven problem structuring method using technology-assisted, recursive component subdivision, leading to a radially hierarchical, visual representation.
- We propose goal-setting as the initial phase (Kanie et al., 2019), prior to problem structuring, and stipulate a singular,

- one- to three-sentence statement of the *superordinate goal* (Sherif, 1958) as the product resulting from this phase before proceeding to problem structuring.
- We propose the construction of a *problem-solving model* in each application of the proposed framework—a visual tool to represent and communicate the stakeholders' work and to keep track of the ongoing progress during later solution implementation.
- To construct the problem-solving model, we define a set of four phases—goal, obstacles, solutions, resources—and emphasize the importance of maintaining focus on one phase at a time, which we call *phase coherence*.

The concept of the visual *problem-solving model* to be constructed by the stakeholders is the anchor of the proposed framework. Its creation is the objective of facilitated stakeholder meetings; its visual, technology-assisted presentation on the big screen—simultaneously accessible by stakeholders through their web portal and mobile app—is used to maintain stakeholder focus throughout the four phases of its construction. Once constructed, it maintains stakeholder focus and displays the status and progress of solution implementations throughout the problem-solving effort. It thus serves as a trusted, enduring, dynamic communication tool that can also be navigable and interactive—if not impactful and immersive—and even publicly accessible.

## 3 Problem-solving Model

This section describes the problem-solving model and the multi-stakeholder process used to create it, starting with our derivation, based on insights provided by the individual, human, cognitive problem-solving process. The reasons for choosing this derivation are three-fold. First, though assisted appropriately by technology, we believe that human insight, trusted relationships, good-will, and perseverance are most likely to result in the *intended consequences* of problem-solving. Second, we believe a problem-solving framework built to reflect the individual human, cognitive problem-solving process will find more resonance with these same human individuals, the actors within the system, to obtain their buy-in, engagement, and motivation. Third, imitating a natural system as a basis for human endeavors has a long history of success, from human flight to architecture to materials science to medicine.

### 3.1 Case of individual cognitive problem-solving process

Condell et al. (2010) define a problem as, "whenever the present situation is different from a desired situation or goal." Robertson (2017, Chapter 1) writes of the individual human problem-solving process in one view as a process of "initial state, solution procedure, goal state," and in another as a process that begins with identifying and defining the problem, then analyzing the data and forming a strategy, organizing information, allocating resources, monitoring progress, and finally, evaluating the outcome.

We propose a symbolic model of this process, where an individual, represented by $i$, perceives a circumstance in their surrounding world, which we will call the system, and the current circumstance, the conditions of the system, as perceived and understood by the individual, which we will call the current or initial system state, labeled $g_0$.

Next, we postulate that the individual imagines a different, desired system state, symbolized by $g_i$. Note that imagining a desired state is not a verbal activity, as neither verbalization nor writing

**Figure 1.** Visual problem-solving models representing an individual i in different scenarios.

(a) The simplest scenario is depicted with the goal state, $g_i$, one obstacle, $o_1$, and its one solution, $s_1$.

(b) A slightly more complicated scenario is depicted, where goal state $g_i$ is hindered by two obstacles—$o_1$ being slightly less impactful than $o_2$—each with its respective solution, $s_1$ and $s_2$.

(c) A scenario as in (b) is depicted, except obstacle $o_2$ has two solutions, $s_{2-1}$ and $s_{2-2}$, the latter having the larger impact on $o_2$.

(d) Obstacle $o_1$ did not have a readily discernable solution, but by subdividing $o_1$ into two component sub-obstacles, $o_{1-1}$ and $o_{1-2}$, solutions $s_{1-1}$ and $s_{1-2}$ became evident.

(e) The scenario in (d) is shown once resources are invited to participate, matching the solutions seeking implementation and the resources' capacities to implement these. Note that solution $s_{1-1}$ has found a match with two different resources, $r_{1-1-1}$ and $r_{1-1-2}$, which cooperate in implementing $s_{1-1}$, the latter contributing proportionally more in its implementation.

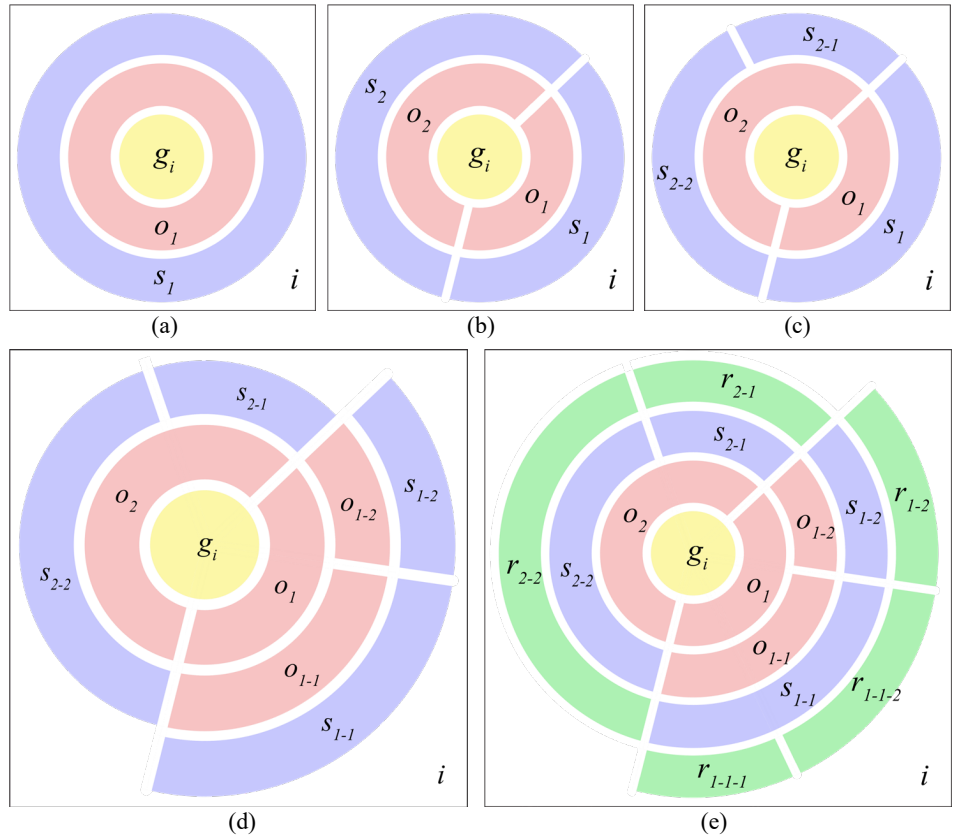

are required; verbalization and subsequent written articulation of $g_i$ will play an important part in our proposed framework.

Rather than represent a path from state $g_0$ to state $g_i$, as per Condell et al. (2010), we propose the following interpretation: The individual identifies the hindrances or obstacles considered to be preventing the current system state $g_0$ from reaching the desired state $g_i$. In our simple case, we represent a single obstacle as $o_1$. We then postulate that the individual will next consider how to overcome obstacle $o_1$ using an imagined solution $s_1$. Finally, the individual will consider what resources are required to implement solution $s_1$, either employing existing resources or determining how new resources can be harnessed.

In the examples given in Figure 1, we label different elements with symbols that will serve the present description; in the case of a real application of this framework, the elements are filled with the corresponding texts as decided upon by the stakeholders and captured in their presence by a meeting facilitation team.

In Figure 1(a), we represent a model of a desired state, $g_i$, a singular obstacle, $o_1$, and the singular solution to that obstacle, $s_1$. This is a symbolic depiction of the inner process of an individual human being. Obviously, the imagination of the desired state, $g_i$, is unlikely to be a circle, as symbolized here, but rather a living, creative, complicated version held in the mind of the individual. Likewise, the obstacle detected by the individual is not a ring around the center circle as depicted here symbolically; rather, it is something quite real and living, envisioned in the mind of the individual. Finally, the solution to overcome the obstacle, as imagined by the individual, is also a living idea that probably does not take the shape of a ring. The symbolic drawing is not meant to assist the individual in the problem-solving process; rather, this model will lend itself to extrapolation in the multi-stakeholder case.

The most important detail captured in this model so far is the identification of distinct phases in the human cognitive problem-solving process.

We postulate that, knowing the current state, the human being first imagines what the desired state is—how things will be when the problem is solved. Envisioning the desired or goal state is not necessarily a step the individual takes deliberately, or even a step taken by the rational mind, as this understanding of the goal state may appear spontaneously without effort in the individual's mind, or may represent a state of being that is felt rather than seen, e.g., driven by bodily function such as hunger, or rooted in the realm of human emotional experience. Ephemeral as this step may be, when extrapolated into the paradigm of multiple individuals, we assert that it should not be overlooked in multi-stakeholder initiatives (MSI).

Second, after establishing a vision of the goal state, the human being identifies the hindrances—what are the obstacles preventing the current state from reaching the desired state? We define these obstacles in such a way that once the obstacles are overcome, we can say that the system has achieved the goal state. We permit a continuous intermediate state of our system, where some obstacles have been overcome, some perhaps only partially, allowing positive progress toward the goal state, even if not yet completely achieved. The goal state arises from within the individual's mind, while the obstacles as perceived by the individual come from the external world.

Obstacles can be decomposed or subdivided into their components, and those further subdivided into their component parts. This subdivision can be repeated until further subdivision becomes unnecessary due to solutions to the given subdivided component becoming apparent. The subdivision of obstacles topologically forms a directed acyclic graph (DAG) with nodes representing obstacles and edges representing the component relationship with parent obstacle nodes. Obstacle nodes with apparent solutions not requiring further subdivision are *leaf nodes* (Danks, 2014, Chapter 3.2).

Third, the human being imagines solutions in the form of concrete actions which result in the given obstacle being rendered no longer an obstacle.

As a final step, the human being matches resources to solutions so that these solutions can be implemented.

### 3.2 Example

Graduate student $i$ is studying at home in the late afternoon, using only the natural light of the sun through the window. As the sun begins to set, the student suddenly is aware of struggling to read in the dim light. The student imagines the goal state, $g_i$, in which he or she continues studying under plentiful light. The student identifies the obstacle, $o_1$, namely, the room lighting is off. The student then identifies the solution, $s_1$, which is to walk to the light switch and turn it on. Now solutions must be matched to resources. In this case, two resources present themselves: the student can stand and walk over to turn on the light switch, or the student's roommate can be asked to do the same. The roommate, already standing and closer to the switch, is glad to turn on the switch, thus implementing solution $s_1$ as resource $r_1$.

### 3.3 Dimensions of model flexibility

The choice of graphical depiction of this model, with goal state, obstacles, solutions, and resources, permits the capture of various dimensions for problem-solving models as they become more complicated, and especially when we translate the model template to the multi-stakeholder case.

The first dimension is the number of obstacles. The model permits any number of obstacles to be identified, labeled $o_1, o_2, \ldots, o_n$, and depicted as semi-circular arcs whose size or *central angle* is proportional to the relative importance of each obstacle, as perceived by individual $i$, to the goal state. In the example in Figure 1(b), two obstacles are depicted, $o_1$ and $o_2$, each with its respective solution, $s_1$ and $s_2$. Obstacle $o_2$ is depicted as playing a slightly larger role than $o_1$ as a hindrance to achieving goal state $g_i$. The next dimension, multiple solutions $s$ for a given obstacle $o$, is depicted in Figure 1(c). In this example, two perceived solutions to obstacle $o_2$, $s_{2\text{-}1}$ and $s_{2\text{-}2}$, are identified by individual $i$, and their size or central angle are depicted in proportion to the individual's perceived impact of the respective solution on the obstacle.

Finally, we wish to represent the possibility that an obstacle $o_1$ may not find a direct solution in the mind of the individual until subdivided into sub-obstacles $o_{1\text{-}1}, o_{1\text{-}2}, \ldots, o_{1\text{-}n}$, where the size of the sub-obstacle arc represents its proportional role—in the mind of the individual—as a component of the parent obstacle. One level of obstacle subdivision is depicted in Figure 1(d), where obstacle $o_1$ is subdivided into $o_{1\text{-}1}$ and $o_{1\text{-}2}$, and the sub-obstacles are associated with solutions $s_{1\text{-}1}$ and $s_{1\text{-}2}$, respectively. Additional levels of obstacle subdivision are also possible. The subdivision of obstacles into sub-obstacles and sub-sub-obstacles is a part of the problem structuring that will become even more useful in the multi-stakeholder case.

## 4 Multi-stakeholder Case

While the model we have developed so far is a representation of an internal human process, we want to use the concepts and the graphical representation to guide and assist multi-stakeholder problem-solving processes. To further develop the problem-solving model from the individual case to the multi-stakeholder case, we must address the following changes:

- Each individual brings his or her own internal problem-solving model to the table. Thus, there is a different goal state $g_i$ for every individual stakeholder $i$ which, strictly speaking, is not the same as the $g_j$ of another stakeholder $j$. We define a process for obtaining a goal state description for the multi-stakeholder group created from the $g_1, \ldots, g_n$ of all individual stakeholders.

- Since within a given individual's frame of reference $i$, every obstacle $o$ is tied, by definition, to the individual $g_i$, and every solution $s$ in turn is tied to the corresponding obstacle $o$, we define a process to obtain a multi-stakeholder problem-solving model by unpacking and reassembling the individual $o$ and $s$.

We begin by addressing two relevant dynamics of multi-stakeholder processes.

### 4.1 Incongruent Goals and Misaligned Phases

Two issues that can make multi-stakeholder efforts seem frustrating and appear unproductive are a lack of *goal congruence* and inadequate *phase coherence*, which we explain in terms of the individual problem-solving model developed in the previous sections.

**Goal congruence** means that a process has been undertaken to establish a goal state, $g$, for the multi-stakeholder group such that each individual stakeholder $i$ is willing to align their internal $g_i$ to the group's goal state, subordinating their individual goal states $g_1, \ldots, g_n$ to that of the group by some combination of

1) finding commonality between the group's goal state and their individual $g_i$;

2) refactoring non-common components of individual goal states into obstacles $o$ that can be satisfactorily expressed in the problem-solving model;

3) conceding to postpone components not addressed in 1) or 2); or

4) willing to not pursue—at least in the given problem-solving effort—those components which cannot be addressed in 1), 2), or 3).

We denominate the goal state of the multi-stakeholder group the *superordinate goal* after well-known work by Sherif et al. (1958, 2017; 1961). We will return, below, to describe the process for clarifying the superordinate goal for a given multi-stakeholder effort.

**Phase coherence** is defined as follows: The individual problem-solving process identified four distinct stages: $g$, $o$, $s$, and $r$. When individuals come together—in the absence of the discipline of informed facilitation—different individuals will speak and contribute from whatever phase of the problem-solving process their individual perspective currently corresponds to. Individuals representing perspectives from the $o$ phase will attempt to persuade the group about what they think the problem is. "I think the problem is" and similar phrases are markers for a perspective centered in the $o$ phase. "What I think we need to do is" is a marker phrase for identifying a contribution stemming from the perspective of the $s$ phase. "Who will do this?" or "How will we pay for this?" are typical phrases from the perspective of the $r$ phase. Rarely, a voice representing the $g$ phase will also be heard: "What I imagine our system looks like once we have resolved the issues is..." All these phases, perspectives, and contributions are valid and valuable and will be welcomed. The point in attaining *phase coherence* is to help the group focus on one phase at a time and to make only contributions that are appropriate to the phase they are focused on at a given time. With disciplined and competent facilitation, the tendency of different phase perspectives to be discussed simultaneously can be guided, postponing the discussion of resources $r$ until solutions $s$ have been identified by the group, postponing the discussion of solutions $s$ until obstacles $o$ have been identified by the group, and postponing the identification of obstacles $o$ until the group has first agreed on the superordinate goal, $g$, i.e., "What does this system look like once we have successfully resolved all issues?"

### 4.2 The Superordinate Goal Phase

The *superordinate goal* is the external representation, in the multi-stakeholder case, of the goal state, $g_i$, described in the internal, individual problem-solving model. It is also the first phase to be undertaken in the facilitated interaction of stakeholders.

#### 4.2.1 Superordinate goal characteristics in the individual case

We describe the following characteristics of the individual goal state, $g_i$, which we will try to reproduce in the multi-stakeholder case.

First, the internal goal state is intuitive; it is held in the mind of the individual.

Second, it requires no language; it is visible and readily accessible—and even navigable and explorable by the individual—without ever being expressed in spoken words or written language.

Third, it represents the individual's balance between their view of the big-picture and the details.

Fourth, though it is reasonable and sensible in the individual's mind, it may be unabashedly and unapologetically ideal, and need not be limited by budgets or even by the laws of physics. Those laws of reality are dealt with in the subsequent problem-solving phases, where obstacles $o$ and solutions $s$ are defined, and resources $r$ are matched.

Fifth, the internal goal state, as imagined by the individual, is not merely a static painting. Though it includes mental still images of the goal state, it also includes mental "video segments" and moving-picture scenarios and even a sense of the relationships between the entities and characters occupying that goal state, what the movements are, how things work in the goal state, how they change over time, and even the emotions of the inhabitants enjoying the goal state.

We set out to find how to externally reproduce these internal goal state characteristics in a way that is practical and useful for multi-stakeholder problem-solving and that engages the individual goal state, $g_i$, and serves as its extension.

### 4.2.2 Superordinate goal requirements in the multi-stakeholder case

For multi-stakeholder problem-solving, we know that the superordinate goal must have a few characteristics.

First, it must be easily expressible by any stakeholder so that it can be communicated, during formulation and development discussions, from one stakeholder to another and understood clearly by the other stakeholders.

Second, it must be adjustable through a facilitated interpersonal process of discussion and negotiation, and practical enough to take place in a group among stakeholders, where all the stakeholders can watch and understand an incremental editing process.

Third, it must be enduring, such that any version of the superordinate goal under discussion, including the finalized expression, lasts beyond the mere discussion in the presence of the stakeholders, so that when they take up work later with the superordinate goal, they will trust it is the same expression they had last been working with.

### 4.2.3 Definition of the superordinate goal

The superordinate goal is a one- to three-sentence written statement. It is a description of the state of our system as we all agree it should look once our problem-solving work is complete, written physically and displayed visually so that all stakeholders can simultaneously view it and participate in the editing process. All stakeholders in the multi-stakeholder effort participate in the editing, which concludes once agreement is reached and all stakeholders can wholeheartedly confirm, "Yes, this is the goal state that I feel we should all work towards." It is not a long, prescriptive, nor descriptive treatise on all the details, as that would disregard many characteristics of the inner goal state we are trying to reproduce. Rather, it is an eloquent, yet short, concise, and artful composition, crafted with the aid of the facilitation team to express the big picture and enough details to be accepted by the stakeholders as a comprehensive description of the goal state. It will appear written in the center circle of our visual model, so it must fit and be legible when viewed by the stakeholders during the stakeholder meetings.

### 4.2.4 Superordinate goal process

Obtaining a successful superordinate goal begins with articulating an initial draft during a multi-stakeholder meeting, projected for visual inspection by all stakeholders, for example via a wall-sized video projector, where the attention of all stakeholders can be directed to the task of changing and editing the statement. The facilitator in this meeting proposes an initial draft statement based on interviews with stakeholders and asks, "What change should I make to this statement so that you feel this statement represents how our system will look once we have finished our work and solved the problems we are here to solve?" Participants suggest changes one at a time, which the facilitator artfully integrates into the given draft, making the appropriate changes visually and in real time. The stakeholders can see the written superordinate goal statement and its editing process visually and can see that all the other stakeholders can also see it.

In giving guidance to the stakeholders concerning the superordinate goal statement, the facilitator may explain, "This statement should represent how our system will look once we are successful in solving all the problems we're here to solve." And, "This statement describes what our system looks like when it is 'fixed,' when it is working as it should."

To help maintain phase coherence during the discussion work—should stakeholders begin to speak about obstacles, solutions, or resources—the facilitator may offer guidance similar to the following. "We will look at obstacles and solutions at a later phase in our process. We won't include any obstacles here, nor any solutions, because we will be extensively and exhaustively covering those aspects later. This moment is for describing the working system, as we all want to see it, assuming we apply all the appropriate solutions and overcome obstacles that we know of or that might arise."

Stakeholders may also inappropriately introduce comments about the resource phase during the "goal state" phase, another example of phase incoherence. A typical facilitator response might be, "We will talk about resources and budgets at a later phase in our process. We won't prematurely limit ourselves now by superficially applying resource and budget constraints, as we will apply these limitations strictly and thoroughly in a later phase. We don't know what new technologies or new alliances or new forces for good may get involved to overcome these limitations, so let's focus on what the system looks like, assuming that resource or budget limitations are met, or that resource and budget limitations may be less daunting than we expected."

In the case where inputs from different stakeholders seem irreconcilable, the key is to separate detail from motivation. What is the irreconcilable detail and what is the motivation for that detail? It is the motivation for the detail that can be expressed in the superordinate goal—multiple motivations if necessary—in a way that the stakeholders expressing conflicting details can understand and agree to each other's motivations without needing to agree on the details of how to realize the goal of those motivations. Such details will come to light in the expression of the obstacles, and the stakeholders can be invited to express those details as obstacles in a later phase of the process. For example, one stakeholder may insist on including "serving vegetarian food" in the superordinate goal, whereas another may prefer the formulation "serving meat." The facilitator will ask each their motivation. "In my ideal system, we want to avoid food that unnecessarily puts a strain on ecological systems," says the vegetarian. "The ideal system serves plenty of protein for health," says the other. Note that while the details are conflicting, the motivations are not. The facilitator proposes, "Serving food that takes into account both nutrition and ecological factors." Both stakeholders agree to this formulation.

Another phenomenon that may require facilitator guidance is when a stakeholder expresses a contribution to the superordinate goal in terms of the absence of what is really an obstacle, i.e., "not $o$." The stakeholder can be invited to express the obstacle $o$ during the "obstacle" phase and to imagine what it will mean to the system to be absent of *o*, so that this can be expressed and included in the superordinate goal.

The challenge is to guide the stakeholders and to iterate patiently over draft versions of a statement until all stakeholders are in agreement. This process is also a socialization process (Kennedy & Widener, 2019). It thus reduces friction and builds trust, as well as reframing to invite a *psychology of possibility* (Langer, 2009).

*4.2.5   Example superordinate goal statements*

For a country: "Make the innovation and start-up eco-system in our country renowned in the world, sustainably."

For an urban education ecosystem: "Our children in these 100 blocks of Harlem will enjoy a school and community ecosystem that will eliminate the education gap between them and successful, suburban students."

**4.3   Obstacles, leaf nodes, and solutions**

Having established the superordinate goal in the initial phase, following our model of individual cognitive problem solving, the next phase is the delineation of obstacles *o*.

In the individual problem-solving case, we arranged for obstacles to be subdivided into sub-obstacles and sub-sub-obstacles, and so forth. Note that with multiple levels of subdivision, only the very last subdivision is associated with a solution. When focusing on each obstacle individually, the question will be, "Is there a straightforward solution or a set of solutions for this obstacle *o*?" If not, the causes of obstacle *o* are explored and *o* is subdivided by the stakeholders into its component causes. Each subdivided obstacle is likewise handled, either recognizing that there are straightforward solutions or by subdividing into subcomponents. This subdivision is much like a tree branch, where each obstacle is a node, and the final node, which requires no further subdivision, is a leaf node.

This mechanism of subdividing—and the visual technology to display it and work with it dynamically with the stakeholders—allows for many levels of subdivision while simultaneously providing easy search, navigation, and visualization. Even problem-solving models with thousands or tens of thousands of leaf nodes can be handled, given a three-dimensional display of the problem-solving model, with the leaf nodes appearing on the outer surface of the solid problem-solving model.

If the superordinate goal phase requires disciplined facilitation to maintain stakeholder focus on the first phase and away from the others, the obstacles phase requires an additional measure of logical thinking and likely extensive preparatory work by the facilitation team, interviewing stakeholders individually and preparing an understanding of the overarching obstacle themes and some of their details.

The instructions from the facilitation team are similar: "Now that you, the stakeholders, have worked together to agree what you want our system to look like once our work is finished, we will now work on developing a logical and comprehensive understanding of what you, the stakeholders, hold to be the obstacles to achieving the goal state you have described. What is keeping us from realizing and experiencing that state today? I will take your inputs and project them on the screen, ordering and organizing them into top-level themes, subthemes of those themes, and so forth."

The shape that will eventually emerge will start with an initial ring of obstacle groups or obstacle themes.

The process focuses on one obstacle at a time. The facilitator and the stakeholders determine if the obstacle presents simple solutions; if not, it is subdivided it into its components.

The stopping criterion for the subdivision of obstacles is the following: "Is this obstacle sufficiently subdivided in order for the stakeholders to easily suggest one or more solutions?" If solutions are readily evident to the stakeholders, the facilitator can identify an obstacle as a "leaf node" ready to begin recording the solution ideas related to that obstacle.

This may be illustrated with an example. Consider one stakeholder group trying to improve their city's educational system. Their superordinate goal is that their children enjoy such a quality education system in their city that their test scores and university acceptance and attendance rates are comparable with the national average. Some of the stakeholders, to the surprise of others who may not have considered it previously, identify "Unsafe neighborhoods" as an obstacle to attaining the goal state of the desired high-quality educational system. When, in turn, the facilitator focuses the stakeholders' attention on this obstacle in order to find solutions or to subdivide it further into components, the stakeholders suggest that among the components is "Crime committed on the streets." In turn, this obstacle is visited by the stakeholders with the guidance of the facilitation team, and the stakeholders decide that this obstacle is still too complicated to imagine simple solutions, so they further subdivide this obstacle. One of the resulting subdivided obstacles is "Crime-prone street conditions." After yet another level of subdivision, one of the components is "Streets too dark at night." At this level of subdivision, once the group visits this obstacle, the stakeholders and the facilitation team agree to mark it as a leaf node, because a set of simple solutions becomes apparent, which are noted as solutions in the visual problem-solving model. The solutions they devise are the following:

1. Install new streetlamps,
2. Replace street lighting,
3. Fix street lighting, and
4. Adjust the automatic switch on/off times of streetlamps so that they start shining earlier, stay on longer, and are off for shorter periods at night.

The obstacle "Streets too dark at night" is a leaf node.

**4.4   Resources**

Once the problem-solving model has been completed by the stakeholders—goal, obstacles, and solutions—the invitation to resources can take place. Two observations can be made about this.

First, the solutions proposed by the stakeholders for the many leaf nodes represent what is essentially a long list of "what needs to be done," where any item on the list, when implemented, will—by design—bring the system closer to the desired goal state. Such a list is an important tool when approaching government and philanthropic entities, and even when reaching out to the community at large.

Second, the engagement with resource entities takes on a different perspective by transferring the choice and knowledge of solutions from the resources to the stakeholders. Instead of a resource entity asserting, "We are funding a particular solution because it is among the solutions that we deem most attractive based on our perspective," or "We are offering to provide a particular service because that is what we do best," the criteria for resource engagement are different. The stakeholders can now approach the resource entities to say, "Here are the solutions we as a community have identified as helpful to achieving the goal of the community. Which of these can you help us implement?"

### 4.5 Implementation Phase

After the problem-solving model has been created, implementation can begin. As part of implementation, each solution that finds a resource to implement it can be instrumented with corresponding performance metrics to measure ongoing progress, which is displayed within the visual problem-solving model. Given a measurement of funding spent and the relative impacts calculated from the DAG, an approximate social return on investment (sROI) can also be deterministically calculated on a per solution and per resource basis.

### 4.6 Realistic example

Figure 2 shows a more realistic example with sample texts. In Figure 2, the goal phase has completed and the obstacle phase has begun. Further progress to the solution and resource phases is not depicted.

The capture of these texts and their placement within the visual problem-solving model in the live presence of the stakeholders is an important aspect of how the proposed framework aims to obtain buy-in from the stakeholders, to build trust, and to break down communication barriers.

### 4.7 Characteristics of the problem-solving model

The representation of the superordinate goal, $g_s$, the radially hierarchical subdivision of obstacles $o$ leading to leaf nodes, their recognized solutions $s$ and the relationship of those with available resources $r$ are what we call the problem-solving model.

The problem-solving model is a roadmap that can be visually depicted and explained to stakeholders and their constituencies; it can live as a reference throughout the problem-solving effort; it can be represented graphically, visually, and artistically; it can be used to visually encode and display the progress of any resource, solution, or obstacle; and it can be used to compute partial impact factors on superordinate goal $g$ of any obstacle $o$, solution $s$, or resource $r$ via the multiplication of proportional impact factors constructed by the stakeholders. Given a measure of financial resources expended by a resource $r$, one can approximate how much any resource $r$ "moves the needle" per unit of funding.

## 5 Assumptions and limitations

We have developed this framework so far by making some implicit assumptions. We would like to make these more explicit, parameterize them, and determine their boundaries—an envelope within which this framework may be applicable.

In general, the intuitive preconditions for this framework to be workable are those described in the following subsections, where we define one problem as having one superordinate goal, $g_s$.

### 5.1 Sufficient stakeholder desire for issue resolution

In order to apply this framework, the stakeholders must desire a resolution to their common problem. A case where the proposed framework may not be applicable or workable would be if there were not sufficient stakeholders who recognize the problem as a problem, or who may recognize the problem but not agree that it needs to be resolved, or who may desire to perpetuate a problem for other purposes. The reasons for the latter may include destructive or even self-destructive motivations, e.g., out of hate or revenge, or due to conditions not conducive to a preponderance of rational thought, such as in the heat of violent conflict.

The application of this framework assumes that the stakeholders agree to participate sincerely in the process. As with any multi-stakeholder framework, achieving stakeholder participation in the process can be challenging.

### 5.2 Appropriate time frames

The time required for the stakeholders to develop the problem-solving model and begin its implementation is presumed to be on the order of weeks to months. The assumption is that the realities of the problem to be resolved will not change considerably during such time frames. Of course, advances in solutions, $s$, and resources, $r$, can occur much more quickly and can be tracked in near real time and reflected in the problem-solving model; even daily or hourly changes are easily reflected within the model. However, changes to the topology and shape of the problem-solving model developed by the stakeholders using the proposed framework should have a much slower time frame. This means that structural changes to the actual problem-solving model—superordinate goal, obstacles, and solutions—should not need to be updated more than about once a year. We define this as $t_r$, the shortest time period between minor revisions of the problem-solving model which the nature and circumstances of the problem can require and still allow the proposed framework to be useful. Major changes or a reformulation of the problem-solving model should be needed only after about a three- or five-year period, which we will call $t_R$, the shortest required time period between major revisions or reformulations of the problem-solving model this framework will allow; a major change is similar to approaching the problem anew.

Changes to the problem-solving model can occur due to internal processes, such as the reduction or resolution of obstacles $o$ thanks to the application of solutions $s$. Changes can also be external, such as natural changes to existing obstacles or the addition of new obstacles that might arise; or a stakeholder change, which can mean the addition of new stakeholders or the exit of others; or changes to the goal state of stakeholders, which can take place over time, as is the prerogative of human stakeholders or the human constituencies they may represent.

The main justification for defining minimum times for minor and major revisions of the problem-solving model—$t_r$ and $t_R$, respectively—is to take into account the need to coordinate the logistics of multi-stakeholder communication and meetings, and the time required for the stakeholders to carry out the processes defined in the proposed framework. The latter process time considers not only the operational and administrative time required, but also the time needed for human stakeholders to reflect, digest, consult with each other and members of their groups, and compose their contributions and responses.

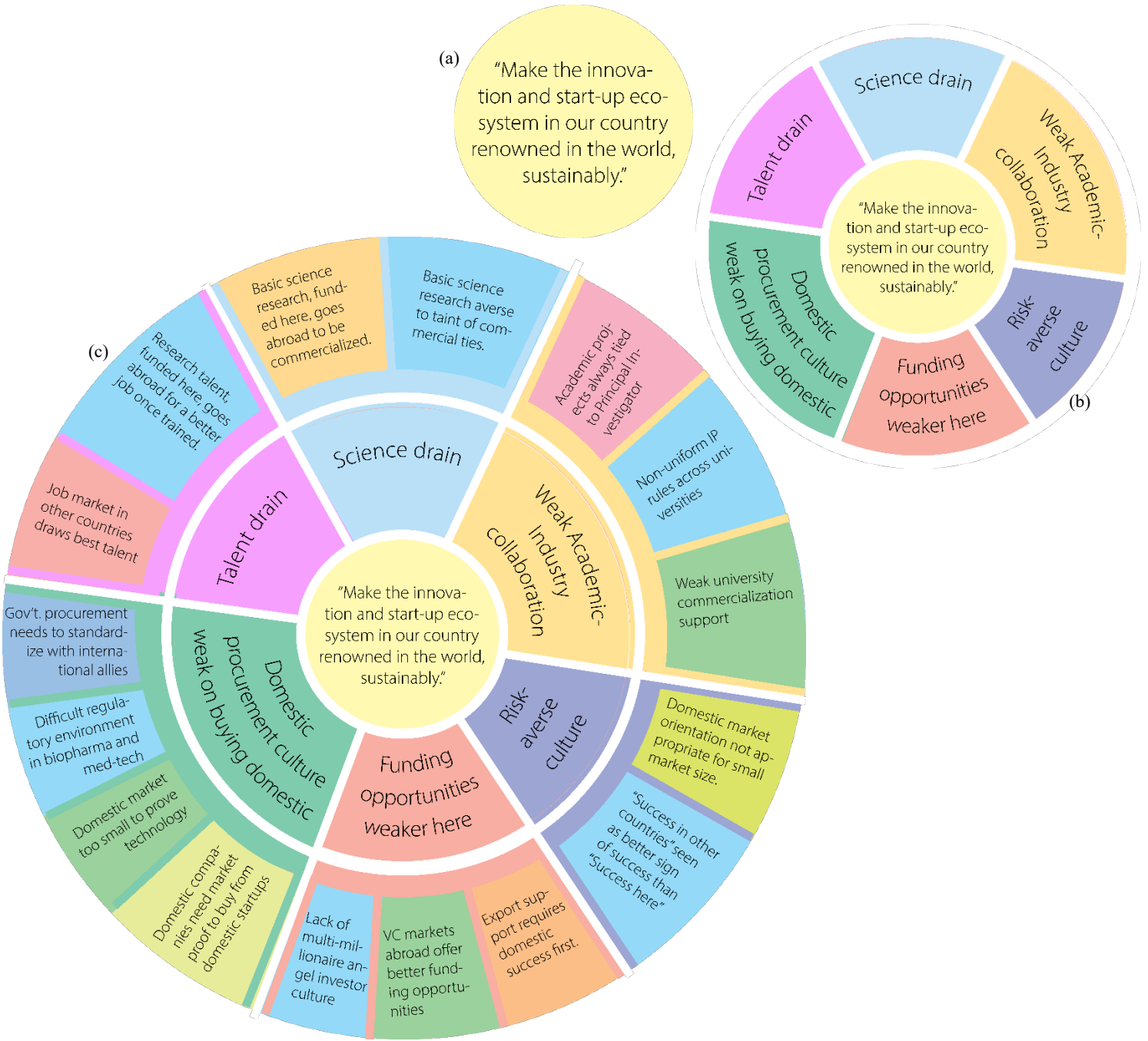

**Figure 2.** Hypothetical example of a problem-solving model in the process of creation in a meeting with stakeholder representatives from industry, government, academia, education, philanthropy, and others who have hypothetically gathered under the proposed framework to address concerns about their country's innovation and start-up ecosystem.

(a) The superordinate goal is formulated by the hypothetical stakeholders. The text is graphically represented.

(b) Hypothetical stakeholders identify six major themes of the obstacles impeding current innovation and the start-up ecosystem in their country from achieving the super-ordinate goal. The themes are captured graphically in the problem-solving model as shown.

(c) Hypothetical example of continued refinement through additional concentric circles where stakeholders agree on the subdivision of obstacles keeping the system—their country—from achieving the superordinate goal they previously articulated. The initial six obstacle themes are subdivided into components. Further refinement by subdivision will continue until the subdivision produces subdivided obstacle components simple enough that solutions become apparent.

This process constructs a radially hierarchical tree, with the superordinate goal at the center, whose leaf nodes will eventually be simple enough obstacles that solutions for these can be proposed. These solutions are also captured in the model. Then, resources are invited to engage and implement them.

Technology that facilitates visually capturing and navigating a very large number of subdivisions of obstacles is used for this process.

## 5.3 Reasonable Complexity

The proposed framework assumes that analysis by the stakeholders—with help from the facilitation team, expert analysis from subject matter experts (SMEs), and the application of data science as needed—will allow a subdivision of obstacles into reasonably independent components. A straightforward subdivision of obstacles is not expected, as such problems are more suited to direct approaches than the proposed framework. Rather, the benefit of the proposed framework is a phased, compartmentalized focus on the obstacles that will help the problem solvers extract components in a radially hierarchical manner to the extent that the problem's realities permit.

In the next section, we consider a measure of a problem's complexity in terms of the formulation in the proposed framework. In other words, hypothesizing that such a formulation is possible, we explore what static and dynamic interactions between elements would be required for a problem to be too complex for the proposed framework.

## 5.4 Parameterizing the Superordinate Goal

We characterize components of the superordinate goal, as agreed upon by the stakeholders, through the lens of individual stakeholder $i$, who has participated in the multi-stakeholder process to determine the superordinate goal, $g_s$, and has accepted its articulation. We formulate an understanding of individual $i$'s imagined goal state $g_i$ prior to participating in the process as comprising various components. First, we recognize the multi-stakeholder superordinate goal $g_s$ as the basis, according to the definition of the process. Then, we allow for a component of the individual's original goal state that is congruent with the superordinate goal, but which may go beyond the superordinate goal in detail or scope while maintaining its congruence. We call this component $g_c$. We then consider a component of the individual's original imagined goal state that was not ultimately reflected in the superordinate goal. We call this component $g_{\bar{c}}$, where $\bar{c}$ indicates non-congruence with the superordinate goal. The stakeholder may have been willing to suppress this expectation within the imagined goal state through a process of compromise, having changed his or her mind or achieved an expanded understanding through the multi-stakeholder process. Alternatively, the stakeholder may have postponed insistence on including this component in the superordinate goal, having decided that the conditions for including it may be more favorable at a later stage, after progress has been made in moving the needle towards the agreed-upon superordinate goal.

We also recognize a component of the individual's imagined goal state which the multi-stakeholders did not include in their agreed-upon superordinate goal statement, but which the stakeholder was able to formulate in terms of obstacles $o$. As part of the goal state definition, these are formulated as the negation of an obstacle, or $\bar{o}$. For instance, the initial individual goal state $g_i$ may include the articulation, "without police corruption." In the resulting problem-solving model, in this example, the superordinate goal $g_s$ does not include the phrasing, "without police corruption," but the problem-solving model does include an obstacle $o$ labeled "police corruption." The stakeholder's original intent to include this as a component in the superordinate goal statement is refactored into a simpler superordinate goal statement plus an obstacle expressed as the negation of that component. We include all such refactoring of multiple obstacles $o$ included in the goal state as a union of the negation of $o$, or $\bigcup \bar{o}$.

The individual's goal state definition is thus expressed as

$$g_i = g_s + g_c + g_{\bar{c}} + \bigcup \bar{o} \qquad (1)$$

This means that the individual stakeholder's goal state $g_i$ consists of four components. One component is the multi-stakeholder superordinate goal $g_s$. The next is $g_c$, that portion of the individual's goal that is congruent with but goes beyond (or falls short of) the superordinate goal, $g_s$. Then, we allow for a component of the individual's goal state $g_{\bar{c}}$ which is not congruent with and cannot be reconciled with the superordinate goal $g_s$. Finally, we consider that an individual may formulate a portion of their goal state in terms of the negation of certain obstacles when formulated in the multi-stakeholder problem-solving model.

Rearranging Eq. (1), we obtain the superordinate goal, $g_s$, in terms of an individual's initial goal state, $g_i$:

$$g_s = g_i - g_c - g_{\bar{c}} - \bigcup \bar{o}. \qquad (2)$$

Thus, the conditions that determine whether an individual stakeholder will be able to agree upon a superordinate goal in the multi-stakeholder process after accounting for the refactoring of negated obstacles, expressed as

$$g_s \approx g_i - \bigcup \bar{o}, \qquad (3)$$

are given by

$$|g_c| \ll |g_s| \text{ and } |g_{\bar{c}}| \ll |g_s|, \qquad (4)$$

where $|x|$ indicates the magnitude, impact, or importance of $x$.

Equations (3) and (4) indicate that the stakeholder is able to agree to the superordinate goal $g_s$ if the strength and impact of $g_c$ and $g_{\bar{c}}$ are sufficiently small compared with the importance of the superordinate goal $g_s$. The stakeholder is able to compromise by finding a way to internally account for components $g_c$ and $g_{\bar{c}}$ by means of postponing, subordinating, or reconsidering components of his or her individual goal state that either go beyond the superordinate goal in detail or scope but are congruent, $g_c$, or are incongruent, $g_{\bar{c}}$, with the superordinate goal.

## 5.5 Parameterizing Complexity of Obstacles, Solutions, and Resources

We now parameterize our assumption of the radially hierarchical subdivision of obstacles $o$, with the aim of finding a usable means of testing applicability of the proposed framework to a given multi-stakeholder socio-ecological system (SES) problem.

The main factor that will impede clean subdivisibility of the obstacles, $o$, is the interdependence of some subdivided obstacles on others—real or perceived—which is problematic in three ways. First, defining the main obstacle themes and determining how they should be subdivided in the work with stakeholders may require reworking and refactoring as additional obstacles and potential or real dependencies to obstacles already registered in the problem-solving model are identified by the stakeholders. Second, even if the stakeholders are able to complete the problem-solving model by discounting some interdependencies, once implementation begins, the actual dependencies may lead to unexpected consequences. For example, the implementation of one solution to an obstacle may increase the magnitude of another obstacle. This may be an inherent, static dependence or a dynamic dependence which arises from the timing with which solutions are implemented. Third, if the dependencies are strong and unaccounted for, complex behavior can result, such as the emergence of additional obstacles caused by the actions guided by the solutions. These are obstacles that had not been considered given a view of obstacles and solutions as independent from each other.

To obtain a theoretical measure of undesirable interdependence, we create a network model and apply a measure of complexity to that model. Note, however, that the nodes in this network model are not the actors in the SES, nor are the edges in the network representative of the interdependent relationships between the actors. Rather, this network is expressed in terms of the critical entities constituting the problem-solving model, namely the obstacle leaf nodes $o$, their solutions $s$, and the corresponding resources $r$ as nodes. The edges of this network represent the interdependence between these entities, and not the actors in the SES system itself. We denominate the constructed dependencies "primary" dependencies, and the undesired dependencies outside of the radially hierarchical model "parasitic dependencies."

For the present analysis, we assume any reasonable, calculable measure $H$ ("Measures of Complexity," 2015; Wackerbauer, Witt, Atmanspacher, Kurths, & Scheingraber, 1994) to compute the complexity of the network,

$$h = H(X_p) \quad (5)$$

—remembering that $X_p$ is not the network of SES actors, but a network consisting of nodes and interdependencies in the problem-solving model—such that we can stipulate a complexity limit $h_{\text{crit}}$ above which the difficulties in formulating a radially hierarchical problem-solving model are too severe, or the approximation of ignoring the parasitic dependencies too inaccurate to use for problem-solving. Thus, the complexity condition for applying the proposed problem framework is stated as

$$h < h_{\text{crit}} \quad (6)$$

Note that the accurate calculation of $h$ depends on the accuracy of $X_p$, such that incorrectly placed or missing nodes and primary dependencies are an additional source of potential inaccuracy of $h$.

## 6  Conclusion

A framework is proposed to provide process and structure to multi-stakeholder initiatives aimed at problem-solving in socio-ecological systems. To model the problem-solving process, a system of visual and symbolic representations is developed, suitable for use in visual presentation tools employed during multi-stakeholder facilitation and later solution implementation efforts. The framework proposes a natural sequence of distinct steps. Emphasis is placed on how countering the tendency for *phase incoherence* in multi-stakeholder participation can help improve focus and productivity in the problem-solving process by avoiding inter-step entanglement.

The individual problem-solving process is modeled as a basis for the multi-stakeholder process, recognizing the superordinate goal as a key to establishing goal congruence. The process of establishing an agreed-upon superordinate goal is emphasized as a distinct exercise to be completed before beginning other phases of the problem-solving process. The benefits include improved trust, an increased sense of promise for a common future vision, and improved positivity by reframing the problem-oriented narratives into a *future-pull* narrative (Land & Jarman, 1992). One practical approach to facilitating this exercise is explained in in Section 4.2.4.

The symbolic and visual model is meant to provide 1) a common theoretical understanding of the problem-solving process for facilitators and stakeholders to reference during the problem-solving process; 2) a logical structure which lends itself, with full stakeholder participation, to technology-assisted creation, exploration, and management of the problem-solving model, a mathematical characterization of the created problem-solving model, and data-driven analysis and management of the solutions in the implementation phase; and 3) a reliable, believable, publicly-accessible representation of the current status of the problem model, which can be impactful, dynamic, and persistent, in order to obtain and maintain stakeholder buy-in throughout the problem-solving effort.

The proposed problem-solving framework aims not to model the behavior of the socio-ecological system in question, but rather to provide a structure and process for stakeholders to logically approach problems when there is a desire and willingness to improve the system, but the question, "Where to begin?" has no clear answer. Within the limitations of parameters, the result is a list of "what needs to be done," constructed by the stakeholders themselves. Resources from government, non-profit organizations, academia, for-profit corporations, and even individual citizens can ask, "What can we do to help?" Any item chosen from the list, by construction, will contribute to moving the needle in the right direction.

No attempt is made to model or predict the actual impact of the solutions on the system, including emergent behavior and unintended consequences. The latter are important phenomena which are taken into account within the proposed framework by adding corresponding obstacles, $o$, in the problem-solving model.

The proposed framework permits the problem-solving model to consider stakeholder input to approximate the relative impacts of obstacles on the superordinate goal, of solutions on obstacles, and of resource efforts on solution implementations. Data science can be important in helping determine the impact coefficients for these inputs. Given measurements of funding and philanthropic inputs to finance efforts by resource entities, this framework would permit a plausible measure of social return on investment. From the point of view of the funding provider, it implies guidance in answering the question, "How can I provide funding to best 'move the needle'?" in a given problem space.

## 7 Acknowledgements

This work was inspired by the philosophy of Value Engineering developed by Lawrence D. Miles (1961)—as applied to large-scale human systems—and a model of human motivation proposed by Graves (1970, 1974), who was a colleague and collaborator of Miles' (Kaufman & Woodhead, 2006, p. 203; Miles, 1961, p. 213). Beck and Cowan (1996) described a theoretical framework including both the problem-solving approach and the human motivation factors, building on the work of Miles and Graves. A case study using the Beck and Cowan framework was published by Beck and Linscott (1991), chronicling their action research while developing this approach in Apartheid-era South Africa. While Value Engineering in manufacturing solves for improved *solution costs* given a desired *function*, Beck and Graves aimed to solve for *functional solutions* given a desired *system goal-state* at an indeterminate cost. The Value Engineering philosophy is the same in both manufacturing systems and human systems: determine first where the system should go—or what the manufactured widget should do—and then work backwards to first identify and subdivide obstacles until functional solutions come into view, and then apply resources.